\documentclass{article}
\usepackage{spconf,amsmath,graphicx}

\usepackage{enumitem}
\setlist{nosep, leftmargin=14pt}

\usepackage{mwe} %
\usepackage{booktabs}
\usepackage{subcaption}
\usepackage{url}
\usepackage{xcolor}
\usepackage{hyperref}
\hypersetup{
    colorlinks,
    linkcolor={blue!50!black},
    citecolor={blue!50!black},
    urlcolor={blue!50!black}
}
\usepackage[capitalise]{cleveref}

\usepackage{float}
\title{Radio-opaque artefacts in digital mammography:\\Automatic detection and analysis of downstream effects}
\name{Amelia Schueppert$^{\star}$ \qquad Ben Glocker$^{\dagger}$ \qquad M\'elanie Roschewitz$^{\dagger}$}

\address{$^{\star}$MIT \\
    $^{\dagger}$Imperial College London
}
\begin{document}
\maketitle
\begin{abstract}
This study investigates the effects of radio-opaque artefacts, such as skin markers, breast implants, and pacemakers, on mammography classification models. After manually annotating 22,012 mammograms from the publicly available EMBED dataset, a robust multi-label artefact detector was developed to identify five distinct artefact types (circular and triangular skin markers, breast implants, support devices and spot compression structures). Subsequent experiments on two clinically relevant tasks $-$ breast density assessment and cancer screening $-$ revealed that these artefacts can significantly affect model performance, alter classification thresholds, and distort output distributions. These findings underscore the importance of accurate automatic artefact detection for developing reliable and robust classification models in digital mammography. To facilitate future research our annotations, code, and model predictions are made publicly available\footnote{Code available at \url{https://github.com/biomedia-mira/mammo-artifacts}}.
\end{abstract}
\begin{keywords}
Artefact detection, Model robustness, Digital mammography, Breast cancer screening
\end{keywords}
\section{Introduction}
\label{sec:intro}

\begin{figure*}
\centering
\begin{subfigure}{0.16\linewidth}
\centering
\includegraphics[width=.9\linewidth]{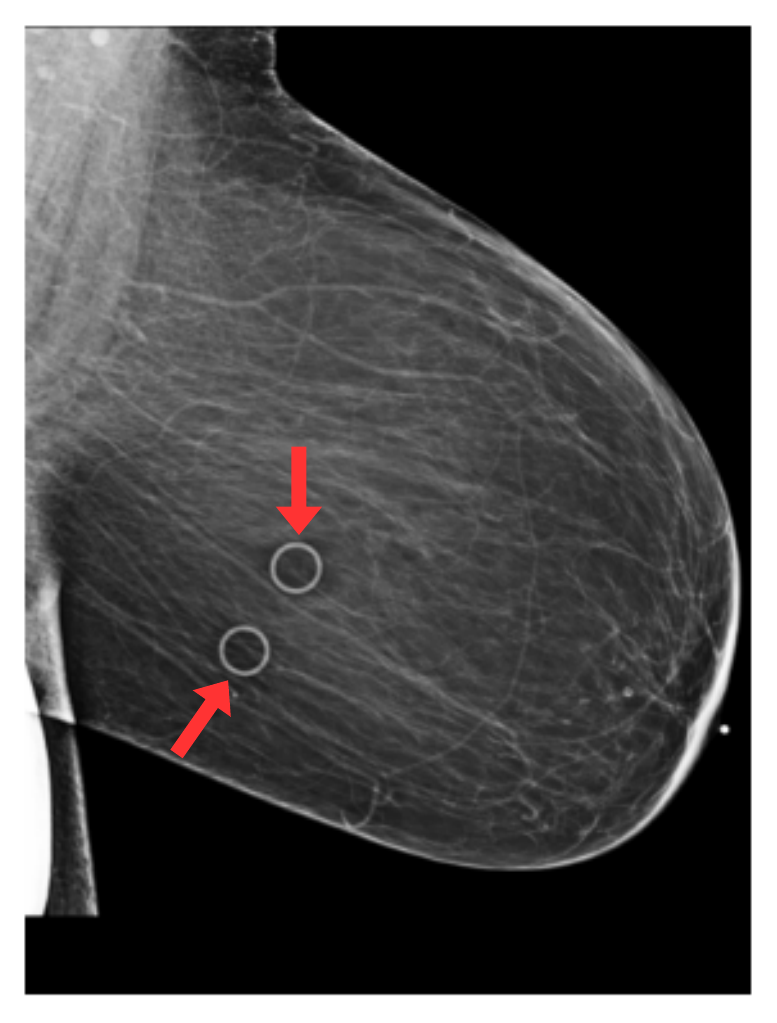}
\caption{Circular marker}
\end{subfigure}
\qquad
\begin{subfigure}{0.16\linewidth}
\centering
\includegraphics[width=.9\linewidth]{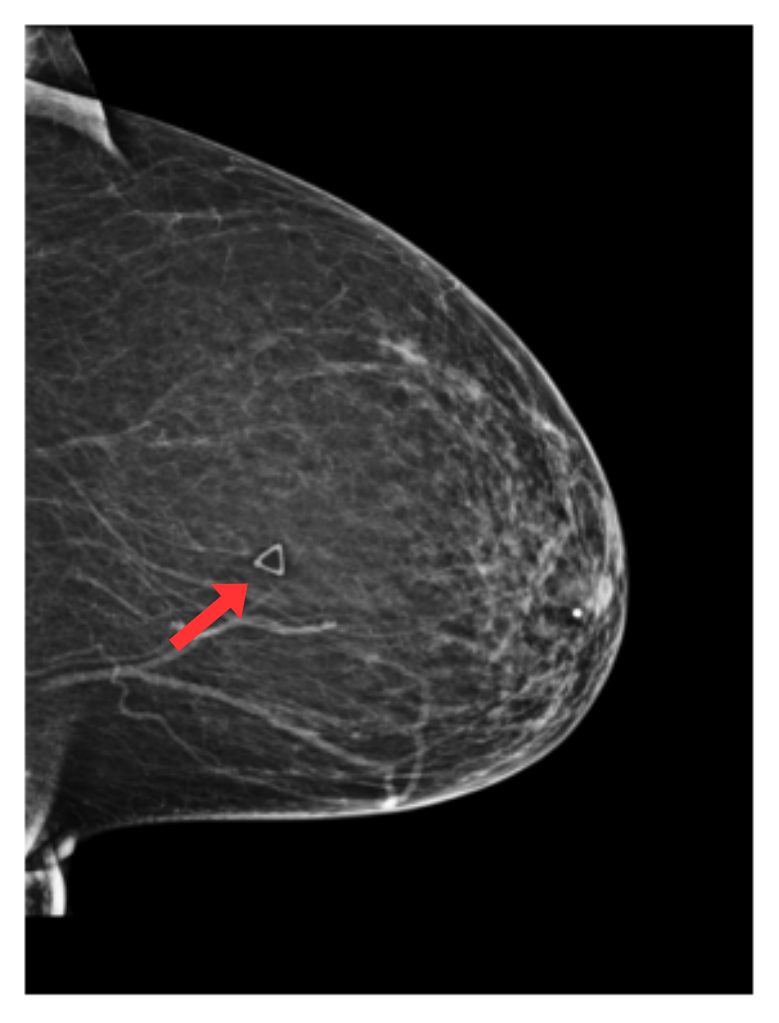}
\caption{Triangular marker}
\end{subfigure}
\qquad
\begin{subfigure}{0.16\linewidth}
\centering
\includegraphics[width=.9\linewidth]{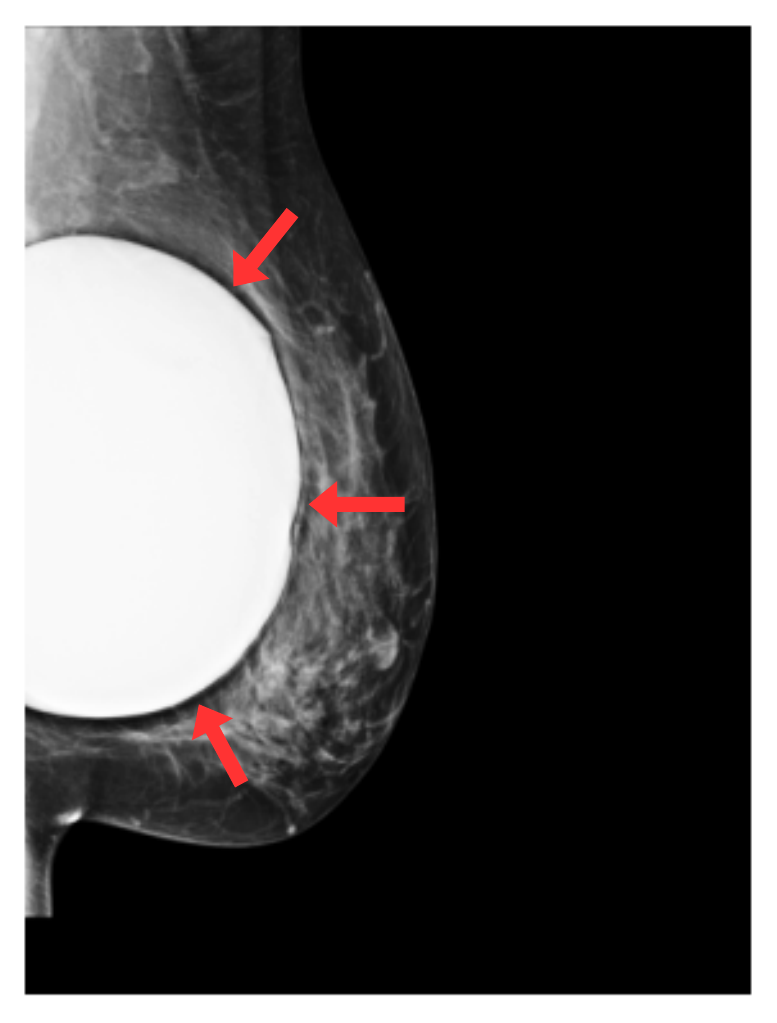}
\caption{Breast implant}
\end{subfigure}
\qquad
\begin{subfigure}{0.16\linewidth}
\centering
\includegraphics[width=.9\linewidth]{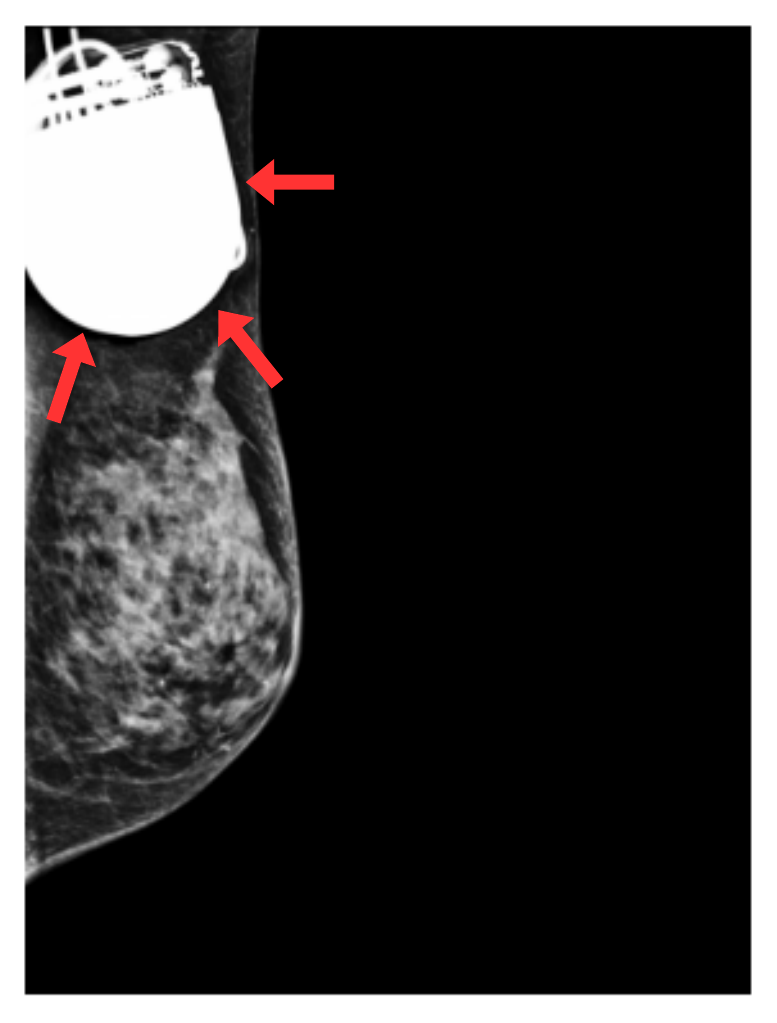}
\caption{Support devices}
\end{subfigure}
\qquad
\begin{subfigure}{0.16\linewidth}
\centering
\includegraphics[width=.9\linewidth]{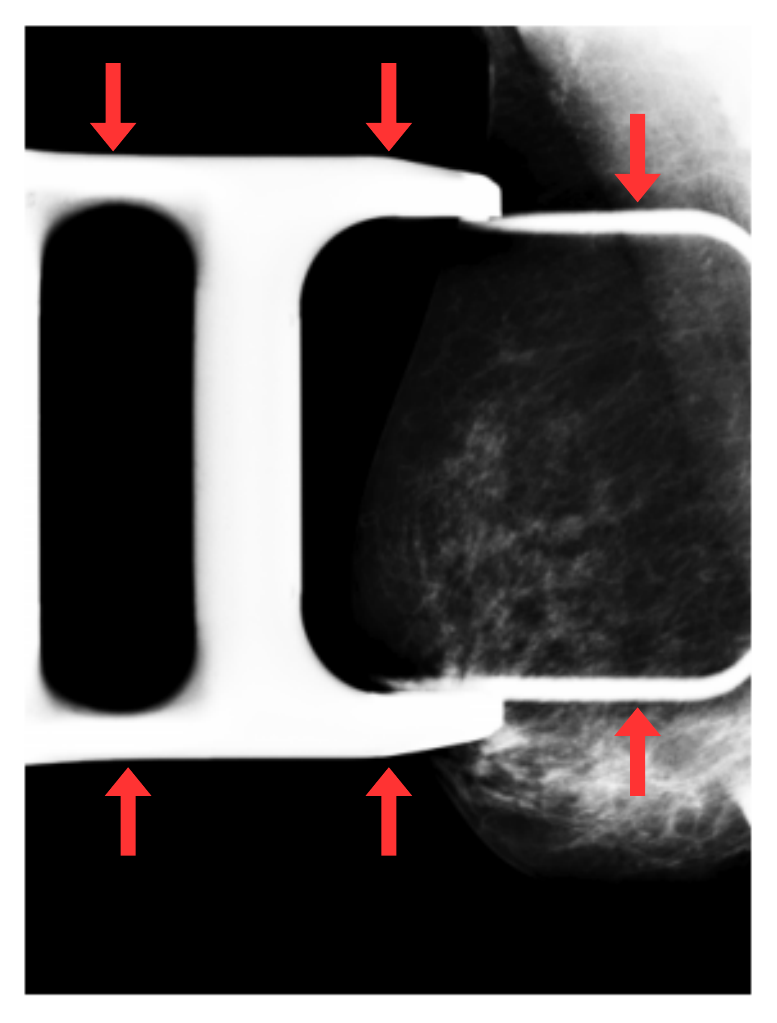}
\caption{Spot compression}
\end{subfigure}
\caption{\textbf{Radio-opaque artefacts considered in this study}: skin markers (circle and triangles), breast implants, support devices (e.g. pacemakers) and spot compression (or magnification) devices. Red arrows highlight artefacts of interest.}
\label{fig:Artefact}
\end{figure*}

The integration of machine learning-based clinical decision tools into breast cancer screening has rapidly gained traction in recent years~\cite{de_vries_impact_2023,dang_impact_2022,lang_identifying_2021,larsen_possible_2022,sharma_multi-vendor_2023,ng_prospective_2023}. These models have demonstrated expert-level performance and the potential to alleviate clinical workload. However, these models have been shown to be sensitive to changes in image characteristics, e.g. scanner changes \cite{sharma_multi-vendor_2023,roschewitz_automatic_2023}. While most studies have investigated the impact of global image changes, this study instead focuses on the effects of radio-opaque artefacts, which introduce localised intensity variations within images.

We focus on five types of artefacts: (a) circular skin markers, (b) triangular skin markers, (c) breast implants, (d) devices (e.g. pacemakers), (e) compression and special magnification artefacts. Examples of such artefacts can be found in \cref{fig:Artefact}. Circular and triangular skin markers are radio-opaque markers placed on the breast skin by radiologists at acquisition time~\cite{ortiz-perez_mammography_2015}, indicating locations of moles (circle) or a palpable mass (triangle). Breast implants and devices such as pacemakers or metallic sensors appear as large, very high contrast structures on the mammograms. Similarly, spot compression and special magnification devices are large radio-opaque structures surrounding the breast area, allowing magnification of small, suspicious regions of breast tissue. Despite their clear effect on image appearances, the effect of the presence of such artefacts in mammograms on downstream task models (e.g. breast cancer screening) has $-$ to the best of our knowledge $-$ not been studied so far. This is largely explained by the fact that labels indicating the presence of such artefacts within an image are not typically available in current publicly available mammography datasets. 

We study this problem in a three step approach, using data from the EMory BrEast imaging Dataset (EMBED), a large publicly available dataset of breast cancer screening mammograms~\cite{jeong_emory_2023}. First, we manually label 22,012 images from this dataset, recording the presence of each artefact type for every annotated image. Secondly, we train and evaluate a multi-label convolutional network to predict the presence of artefacts on the remaining images. Thirdly, we train two downstream task classifiers, the first one for breast cancer screening and the second one for breast density assessment. We then study the effect of individual artefacts on model performance and calibration across both tasks.

Our contributions can be summarised as follows:
\begin{itemize}
    \item We manually annotate 22,012 images of the EMBED dataset, indicating the presence of five artefact types (circular and triangular skin markers, breast implants, support devices and compression structures). All annotations are made publicly available.
    \item We train a multi-label artefact detector able to predict the presence of individual artefacts with over .99 average ROC-AUC, allowing us to predict presence of artefacts for all 398,458 EMBED images.
    \item We analyse the effect of artefacts on downstream model performance for breast cancer screening and breast density assessment. Our results demonstrate that the presence of artefacts can strongly affect downstream performance.
    \item To facilitate future research, all annotations, code and model predictions are made publicly available.
\end{itemize}

\section{Methods}
\label{sec:methods}
\subsection{Artefact dataset construction}
We use the publicly available EMBED dataset~\cite{jeong_emory_2023} which is composed of a total of 398,458 mammograms, and contains full-field digital mammograms (FFDM) as well as so-called C-view images (2D projections obtained from breast tomosynthesis images). Images in the dataset were acquired on six different scanners, covering a wide-range of image characteristics. This dataset does originally not contain any indication of the presence of skin markers, breast implants or devices. Spot compression tags are available as part of the original release. However, our initial analysis showed that this provided tag is unreliable, containing many false negatives. A more accurate detection is required to flag all images with special magnification and compression structures.

We manually labelled artefact presence in 22,012 images from the EMBED dataset. In this annotated dataset: 23\% of images have circular skin markers, 5\% have triangular skin markers, 8\% have breast implants, 1\% have support devices (such as pacemakers or metallic sensors) and 6\% of images contain compression structures. Detailed dataset statistics can be found in \cref{tab:annotations}. We initially labelled a random subset of the dataset, which we then expanded in an active learning manner, using model predictions to guide selection of additional positive and hard negatives samples for each class.

\begin{table}[H]
    \centering
    \begin{tabular}{lc}
    \toprule
    Artefact & Images with artefact \\
    \midrule
    Circles & 4,905 (22\%) \\
    Triangles & 1,186 (5\%) \\
    Implants & 1,815 (8\%) \\
    Devices & 286 (1\%) \\
    Spot compressions & 1,250 (6\%) \\
    \midrule
    Total images & 22,012 \\
     \bottomrule
\end{tabular}
    \caption{\textbf{Characteristics of labelled artefact dataset}.}
    \label{tab:annotations}
\end{table}

\subsection{Artefact detection model}
After creating the large, annotated artefact dataset, we trained a multi-label classifier to predict the presence of such artefact on the remainder of the EMBED dataset. Specifically, we used a ResNet-50 model~\cite{he_deep_2016}, initialised with ImageNet weights. We use a separate binary classification head for each artefact, and the encoder is shared across artefact types. We used 65\% of the annotated dataset for training, 15\% for validation, and 25\% for testing.

\subsection{Downstream tasks}
To analyse the effects of artefacts on downstream model performance, we focus on two clinically relevant tasks. First, breast cancer screening, where we separate normal images from mammograms with non-negative screening results. Precisely, a mammogram is said to be `normal' if the BIRADS level is 1 (negative class). If the mammogram is labelled as BIRADS $>$ 1 (i.e. benign, suspicious or biopsy-proven cancer) or if there is any pathology result confirming the presence of cancer for this image, the image is considered non-negative (positive class). Our second task is breast density assessment (4 density classes), a task of interest in breast cancer screening as breast density is a risk factor for breast cancer. Both downstream classifiers were trained independently, using ResNet-18 models, initialised with ImageNet weights. For these downstream tasks, we focused on images with valid breast density assessment, female subjects, and excluded spot compression and C-view images. We used 67.5\% of the data for training,, 7.5\% for validation, and 25\% for testing.

\section{Results}
\label{sec:results}
\subsection{Artefact detector performance}

\begin{figure}
\centering
\includegraphics[width=.95\linewidth]{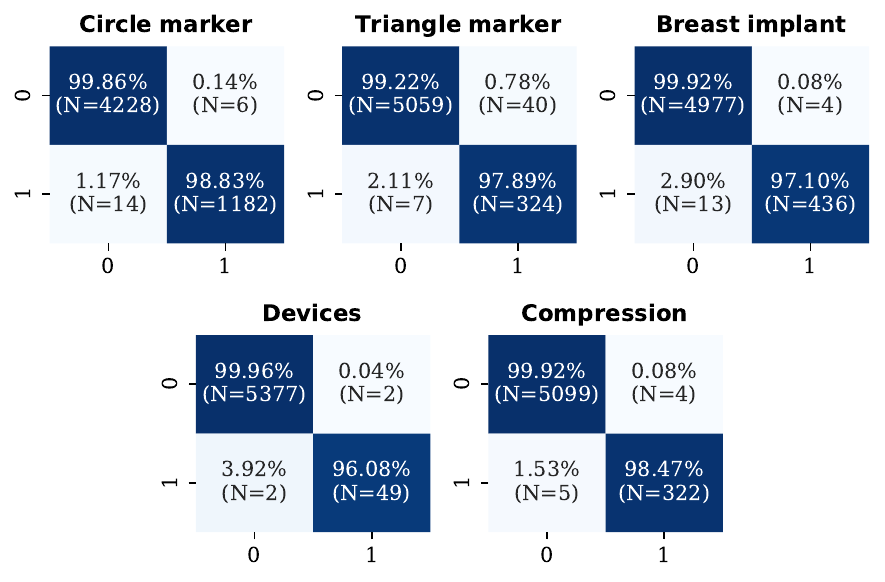}
\caption{\textbf{Artefact detection performance}: excellent detection performance (balanced accuracy $>$ 98\%) for all artefacts.}
\label{fig:confusion_artifacts}
\end{figure}

We first evaluate the performance of the proposed artefact detector on the labelled artefact test set. The model demonstrates excellent performance on these artefact detection tasks, achieving an average ROC-AUC of .993 on the test set and near perfect label-wise detection accuracy, as shown in~\cref{fig:confusion_artifacts}, with a balanced detection accuracy $>$ 98\% for all artefact types.

\subsection{Artefact distribution on the full EMBED dataset}
The excellent performance of the artefact detector allows to confidently predict the artefact tags for the remaining of the EMBED dataset (N=398,458). Overall, the model predictions suggest that 22\% of the images have a least one artefact, indicating that artefacts affect a substantial portion of the dataset, further motivating the need to analyse their effect on downstream model performance. In \cref{tab:predictions}, we compare the distribution of artefacts from images with normal screening results versus images with non-negative findings. We observe noteworthy correlation between cancer screening result and the presence of triangular skin markers, which is expected as triangle markers are associated with palpable masses and suspicious areas. Unsurprisingly, nearly all images with spot compression are associated with a non-negative screening result, as spot compression is typically applied to further investigate potentially suspicious areas. Finally, our model found $756$ additional images with compression compared to the original spot compression tag provided with the dataset.

\begin{table}
    \centering
    \begin{tabular}{lcc}
    \toprule
     & Normal images & Non-negative images \\
    \midrule
    No artefacts & 251,805 (84\%) & 57,699 (59.5\%) \\
    Circles & 31,698 (10\%) & 7,148 (7\%) \\
    Triangles & 3,769 (1\%) & 7,477 (8\%)    \\
    Implants & 8,973 (3\%) & 2,111 (2\%) \\
    Devices & 775 (0.3\%) & 309 (0.3\%) \\
    Compressions & 6,734 (2\%) & 27,370 (28.2\%) \\
    \midrule
    Total images & 301,416 & 97,042 \\
     \bottomrule
\end{tabular}
    \caption{\textbf{Number of images with artefacts on the full EMBED dataset, by cancer screening status, as predicted by our artefact detector}. There are significantly more images with triangles markers and spot compression among images with non-negative screening results.}
    \label{tab:predictions}
\end{table}

\subsection{The effect of artefacts on downstream prediction tasks: breast cancer screening and density assessment}
\begin{figure}
\centering
\includegraphics[width=.95\linewidth]{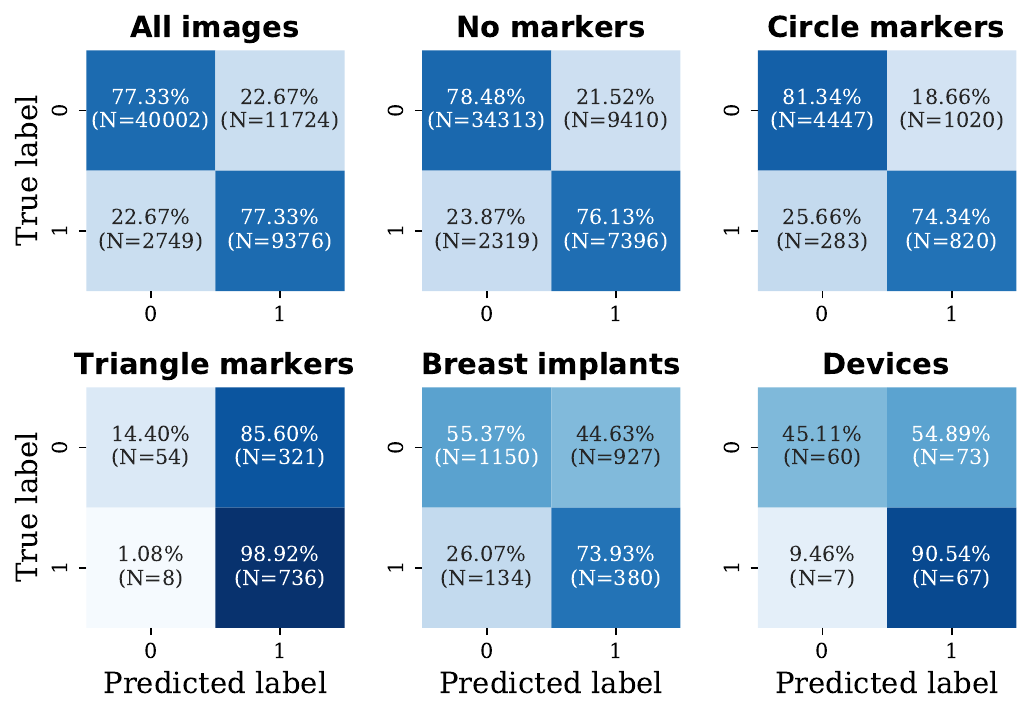}
\caption{\textbf{Confusion matrices for breast cancer screening, per marker subgroup}. The classification threshold is fixed across all subgroups, chosen such that the global sensitivity and specificity are equal. The sensitivity-specificity balance on images with triangle markers, breast implants and support devices is substantially degraded, where images with those artefacts have a very low specificity compared to images without markers.}
\label{fig:confusion_classification}
\end{figure}

\begin{figure}
\centering
\begin{subfigure}{\linewidth}
    \includegraphics[width=.99\linewidth]{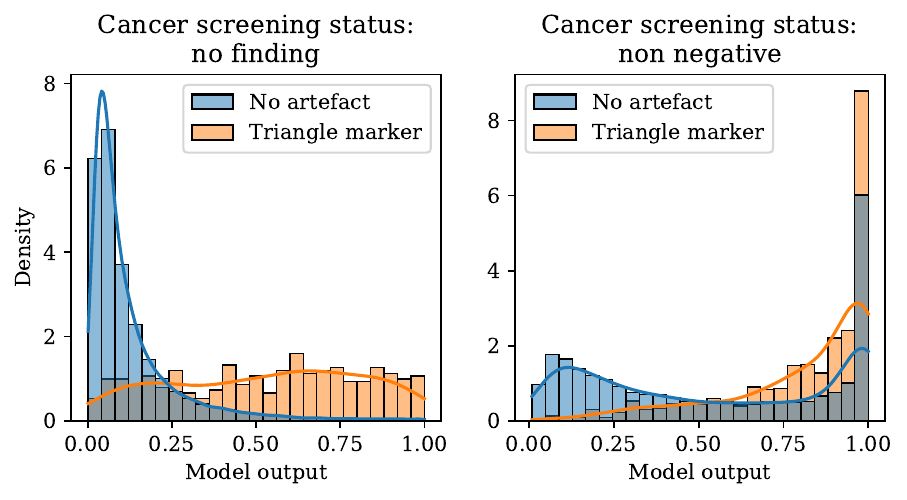}
\end{subfigure}
\begin{subfigure}{\linewidth}
    \includegraphics[width=.99\linewidth]{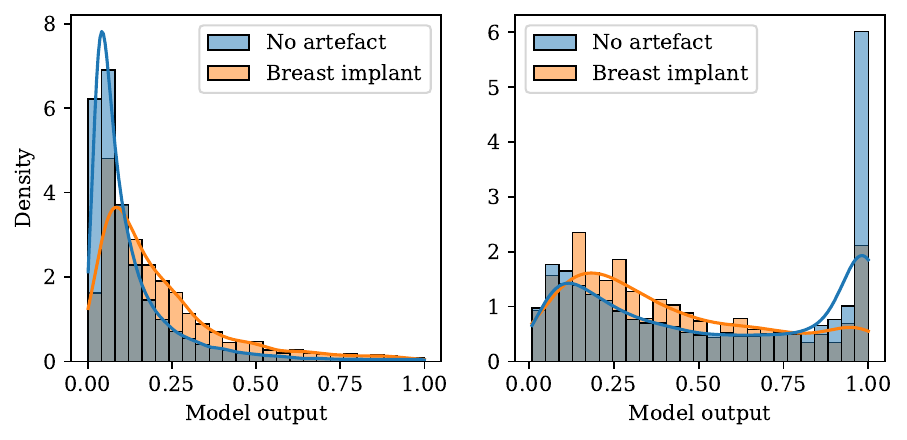}
\end{subfigure}
\begin{subfigure}{\linewidth}
    \includegraphics[width=.99\linewidth]{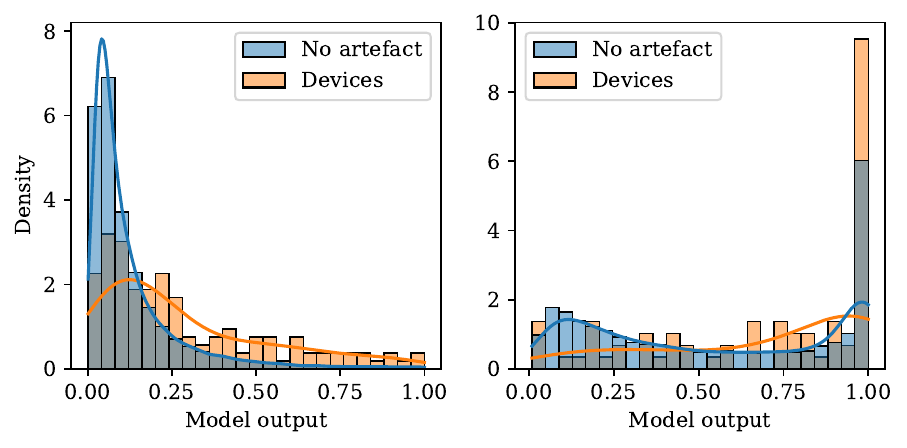}
\end{subfigure}
    \caption{\textbf{Effect of artefacts on model output distribution for the breast cancer screening model}. From top to bottom, effect of: triangular skin markers, breast implants and devices.}
    \label{fig:distribution}
\end{figure}

\begin{figure}
\centering
\includegraphics[width=\linewidth]{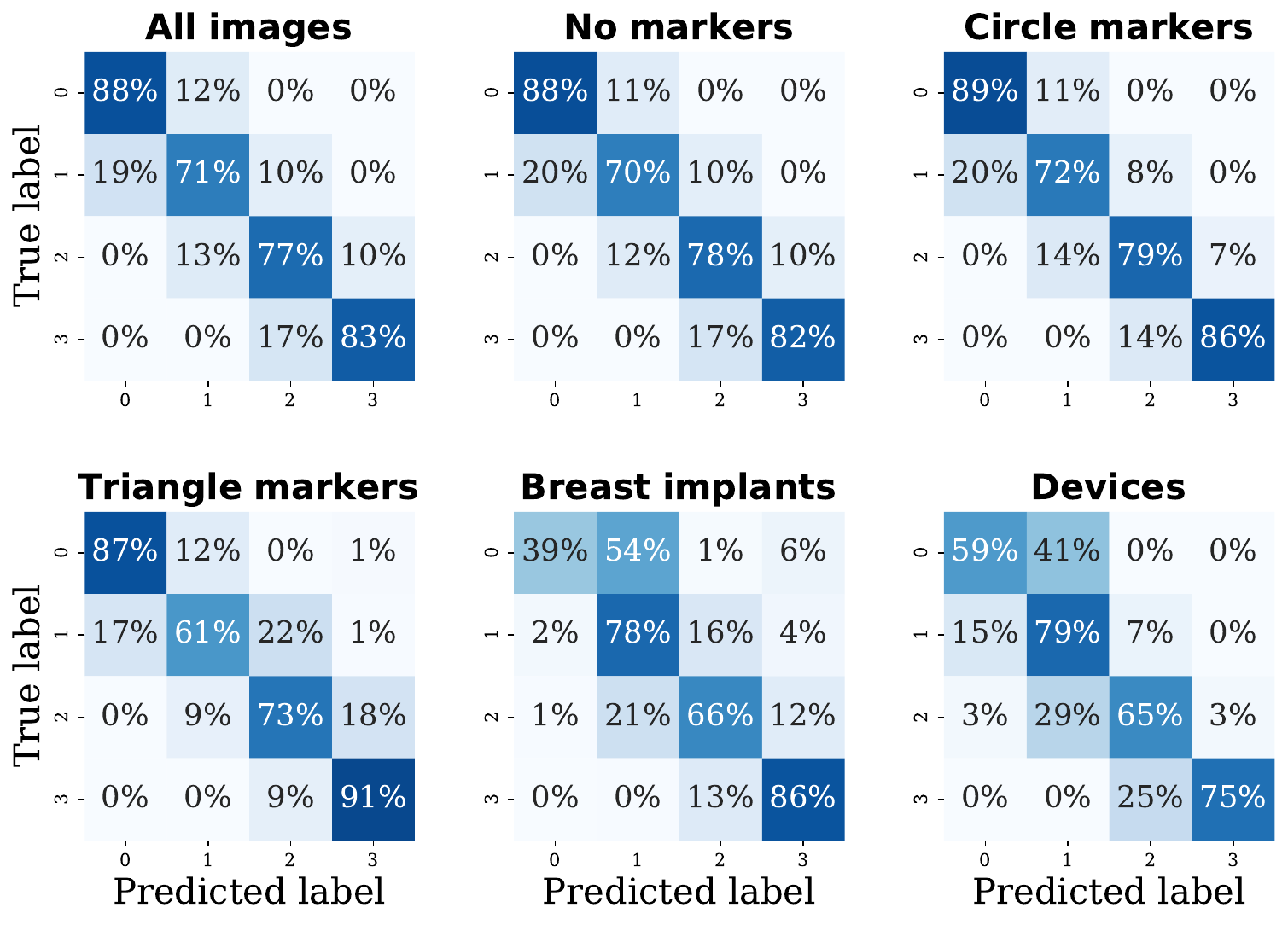}
\caption{\textbf{Confusion matrices for breast density classification, per marker subgroup}. Class-wise accuracies are substantially shifted on images with breast implants and devices.}
\label{fig:confusion_density}
\end{figure}

\begin{table}[b]
\centering
\begin{tabular}{lcc}
\toprule
 & Cancer screening & Breast density \\
 & (ROC-AUC) & (Balanced acc.)\\
\midrule
Overall &  .86 & .80 \\
\midrule
No artefacts & .86 & .80 \\
Circles & .86 & .82 \\
Triangles & .79 & .78 \\
Implants & .72 & .67 \\
Devices &  .83 & .70 \\
\bottomrule
\end{tabular}
\caption{\textbf{Downstream classification performance across artefacts subgroups}. The performance on images with triangle markers, implant and support devices is significantly lower than on images without artefacts for both tasks.
}
\label{tab:downstream}
\end{table}
In this section, we analyse the effect of artefact presence on classification performance for downstream tasks. In \cref{tab:downstream}, we report the overall test performance along with the performance on images where particular artefacts are present, for both tasks (normal versus non-negative breast cancer screening and breast density assessment). We observe that the cancer screening model significantly under-performs on images with triangular skin markers, breast implants and support devices. A similar effect is observed for breast implants and support devices for the breast density assessment model. Next, we analyse the effect of subgroups on the distribution of model outputs and threshold choice for the binary breast cancer screening task. In \cref{fig:distribution}, we show the distribution of model outputs by ground truth for images without any markers versus for images with triangular skin markers, breast implants and devices. The presence of artefacts induces a clear shift in the model output distribution, an effect similar to what has been previously reported for acquisition changes~\cite{roschewitz_automatic_2023,de_vries_impact_2023}. This shift in model output will in particular affect threshold selection for binary classification. We illustrate this in \cref{fig:confusion_classification}. In this experiment, we fix the classification threshold such that the overall sensitivity and specificity are equal on the overall test set, and then report the subgroup-wise confusion matrices for various artefact subgroups. We can see that images with triangle markers are systematically predicted as positive, yielding an extremely low specificity on this subgroup. Similarly, we can observe substantial drop in model specificity for images with breast implants and support devices. Similar effects can be observed for breast density assessment in~\cref{fig:confusion_density}, where class-wise accuracies vary substantially.

\section{Conclusion}
\label{sec:conclusion}

 In this paper, we developed and validated a multi-label classification model for detecting artefacts in digital mammography images. We manually annotated over twenty thousand mammograms for the presence of five different types of radio-opaque artefacts (circular and triangular skin markers, breast implants, devices and compression/magnification structures). These annotations are made publicly available, along with the developed labelling tools to facilitate further research in this space. We additionally release the generated annotations for the entire EMBED dataset (398,458 images). 
 
 Our experiments clearly show that these artefacts present a challenge for mammography-based machine learning models. We notably showed that the presence of some artefacts may lead to substantial performance drops, changes in output distribution and affect the classification threshold choice.

Accurately identifying and accounting for artefacts in both model evaluation and training is key to uncover important model biases. Our experiments show that care should be taken against the presence of radio-opaque artefacts in mammography datasets and is calling for more research to enhance the robustness of existing models against these artefacts. We believe that the released artefact annotations and artefact detector constitute a important step in enabling further research in this space.

\section{Compliance with ethical standards}
\label{sec:ethics}
This study uses secondary, fully anonymised data which is publicly available and is exempt from ethical approval.

\section{Acknowledgments}
\label{sec:acknowledgments}
A.S. received support from MISTI at MIT and the International Research Opportunities Programme (IROP) at Imperial College London. B.G. received funding from the Royal Academy of Engineering as part of his Kheiron/RAEng Research Chair in Safe Deployment of Medical Imaging AI. M.R. is funded through an Imperial College London President's PhD Scholarship.

\bibliographystyle{IEEEbib}
\bibliography{references_all}

\end{document}